\documentstyle[aas2pp4]{article}
\newcommand{\rsf}{r}
\newcommand{\di}{\varpi}
\newcommand{\MS}{M_{\sun}}

\begin{document}

\title{Old Isolated Accreting Neutron Stars: The Diffuse X--ray 
Emission From The Galactic Center}
\author{Silvia Zane}
\affil{International School for Advanced Studies, \\ Via Beirut 2-4, 34014
Trieste, Italy \\ e--mail: zane@sissa.it}
\author{Roberto Turolla}
\affil{Dept. of Physics, University of Padova, \\ Via Marzolo 8, 35131 Padova, 
Italy \\ e--mail: turolla@astvs1.pd.infn.it}
\and
\author{Aldo Treves} 
\affil{International School for Advanced Studies, \\ Via Beirut 2-4, 34014
Trieste, Italy \\ e--mail: treves@astmiu.uni.mi.astro.it}

\begin{abstract} 
The contribution of weakly--magnetized ($B\sim 10^9$ G) neutron stars 
accreting the interstellar 
medium to the diffuse X--ray emission observed in the Galactic Center 
is investigated. It is shown that, under rather conservative assumptions about
the neutron stars and gas distributions, the accretion luminosity can 
account for a sizable fraction, possibly most, of the detected X--ray flux
in the 2.5--7 keV band. In particular, model results are compared with {\it
Granat\/} data and show a general agreement in both the flux energy
and radial distributions.
\end{abstract}


\section{Introduction}

It was suggested long ago in a seminal paper by \markcite{ors70} 
Ostriker, Rees, \& Silk (1970) that old isolated neutron stars (ONSs) can be 
roused from their 
lethargy by the accretion of the interstellar medium, revealing themselves as
weak, soft X--ray sources. 
About 20 years later it was realized that the capabilities of the 
{\it ROSAT\/} satellite were unprecedented in matching the challenge of 
observing 
ONSs (\markcite{tc91}Treves \& Colpi 1991; \markcite{bm93} Blaes \& Madau 
1993; \markcite{mb94} Madau \& Blaes 1994; \markcite{zan96}
Zane {\it et al.\/} 1996). The basic criteria for 
discriminating ONSs from other classes of X--ray sources should be their
very low luminosity, extreme X--ray to optical flux ratio and
spectral softness. In accordance with the pioneering suggestion by 
\markcite{shv71}Shvartsman (1971), giants molecular clouds were 
recognized as very favourable sites for the detection of accreting ONSs 
(\markcite{bm93}Blaes \& Madau 1993; \markcite{cct93}Colpi, Campana, \& Treves 
1993). 

Despite the increasing efforts, up to now only two 
X--ray sources have been proposed as ONSs accreting the interstellar medium:
MS 0317.7-6647 (\markcite{sto95}Stocke {\it et al.\/} 1995) and 
RXJ185635-3754 (\markcite{wal96}Walter, Wolk, \& Neuh\"auser 1996). In both 
cases the arguments in favour of their identification 
are strong but not completely compelling. 
Systematic searches on available X--ray data are under way
(see e.g. \markcite{dkh94}Danner, Kulkarni, \& Hasinger 1994; 
\markcite{bzc96}Belloni, 
Zampieri, \& Campana 1996), but the positive detection of ONSs, which could be 
as many as $10^9$ in the Galaxy, appears still a difficult goal to achieve.

Because of the intrinsic difficulties in detecting single objects
and of the expected relative abundance of ONSs in the Galaxy, it is natural 
to consider their overall emission and more specifically their contribution 
to the soft X--ray background, as already suggested by 
\markcite{ors70}Ostiker, Rees, \& Silk (1970). 
Assuming $B\sim 10^9$ G and $N_{ONS}\sim 10^9$, 
(\markcite{zan95}Zane {\it et al.\/} 1995) have recently shown that ONSs 
accreting the interstellar gas can account for a non--negligible fraction of 
the observed excess in the soft X--ray background which is not 
extragalactic in origin but could be produced by a Galactic population 
\markcite{mg95}(Maoz \& Grindlay 1995).

Since the diffuse emissivity of accreting ONSs depends on their 
spatial concentration and on the density of the interstellar medium, it is of 
particular interest to consider the Galactic Center, a site where both the 
stellar and gas densities exceed by orders of magnitude those of other regions
of the Galaxy. At the same time the Galactic Center is a well known source of 
diffuse X--ray emission, first detected by {\it Uhuru\/} 
(\markcite{kel71}Kellogg {\it et al.\/} 1971)
and then studied by virtually all X--ray missions. The possible association
of the Galactic Center diffuse X--ray source with accreting ONSs was originally 
suggested by \markcite{mtt73}Maraschi, Treves, \& Tarenghi (1973).

In this paper first we review the X--ray observations of the Galactic Center. 
We then 
discuss the interstellar medium, the stellar 
and the expected neutron star distributions. The calculation 
of the emission due to accreting ONSs is presented and our results are
compared with observational data. Discussion follows.

\section{The Galactic Center}

\subsection{The X--ray Emission}

The Galactic Center is one of the more widely explored regions of the sky 
in the X--rays and observational efforts appear indeed motivated in the 
light of the complexity of the source. The inner 100 pc of the Galaxy 
exhibit, in
fact, a region of diffuse emission together with a number of
point--like sources. The strong absorption in the direction of the Galactic
Center makes its appearance substantially dependent on the energy band
(soft or medium X--rays) in which the source is observed. Here we briefly
outline the current status of X--ray observations of the Galactic Center,
focussing our attention on the diffuse component.

The presence of a weak, diffuse emission from about $\sim 1$ deg$^2$ was 
already suspected in the {\it Uhuru\/} 2--10 keV data (\markcite{kel71}Kellogg 
{\it et al.\/} 1971) and first confirmed by {\it Einstein\/}
in the 0.5--4 keV band (\markcite{wat81}Watson {\it et al.\/} 1981). 
Observations with 10$^\prime$ resolution in the 2--15 keV range, performed 
with {\it Spartan 1\/} (\markcite{kaw88}Kawai {\it et al.\/} 1988) and 
{\it Spacelab 2\/} (\markcite{ski87}Skinner {\it et al.\/} 1987; 
\markcite{ski89}Skinner 1989), confirmed the 
diffuse emission at higher energies. Spectral measurements by {\it Ginga\/},
with an angular resolution of $\sim 1^\circ$, have shown the presence of
a strong emission line at 6.7 keV, which was identified with the K$\alpha$
line of He--like iron (\markcite{koy89}Koyama {\it et al.\/} 1989; 
\markcite{yam90}Yamauchi {\it et al.\/} 
1990). The continuum appears rather flat and is well fitted by a thermal 
bremsstrahlung. A temperature of $12.7\pm 0.4$ keV was calculated
on the basis of TTM observations by \markcite{not93}Nottingham {\it et al.\/} 
(1993).
Such temperature, however, is more than one order of magnitude too high 
for a plasma to be confined in the Galactic Centre potential well.

The ART--P telescope on board {\it Granat\/} observed the Galactic Center
region with 5$^\prime$ resolution in the 2.5--30 keV band with rather long exposure 
times (\markcite{smp93}Sunyaev, Markevitch, \& Pavlinsky 1993; 
\markcite{msp93}Markevitch, Sunyaev, \& Pavlinsky 1993). 
The intensity profiles obtained in 4 energy bands, after subtracting 
point--like sources, show the presence of an elliptical, extended source, 
as first suggested by Kawai {\it et al.\/} Spectra from
the total $1^\circ \times 1.5^\circ$ ellipse and from a central region
30$^\prime$ wide were produced, confirming a flat bremsstrahlung--like 
continuum; a strong absorption feature at 8--11 keV was also detected. 
\markcite{smp93}Sunyaev, Markevitch,
\& Pavlinsky (1993) noted that the structure of the diffuse emission differs 
substantially
below and above $\sim 8$ keV. While the lower energy component is thermal and 
roughly elliptical, the hard emission comes from an elongated region, 
parallel to the Galactic plane, that resembles the distribution of Giant
Molecular Clouds. This led \markcite{smp93}Sunyaev, Markevitch, \& Pavlinsky 
to the conclusion that
the diffuse emission consists of two components and that the high energy
portion of the spectrum may be due to Thomson scattering of hard photons
on the dense material of the clouds. In this picture the bremsstrahlung 
temperature could be lower than the previous estimate 
(\markcite{msp93}Markevitch, Sunyaev, \& Pavlinsky give $T_{brems}\sim 3$ keV), 
easing the problem of confining the hot gas in the Galactic Center.

The best resolution X--ray map of the Galactic Center was obtained with 
{\it ROSAT\/} in the 0.8--2 keV range (\markcite{pt94}Predehl \& Tr\"umper 
1994). In order to 
explain the lack of X--ray sources at the position of Sgr A$^*$,
an interstellar absorption higher than $2\times  10^{23}$ cm$^{-2}$
was invoked by \markcite{pt94}Predehl \& Tr\"umper. Preliminary reports of ASCA
observations (\markcite{koy96}Koyama {\it et al.\/} 1996) indicate the 
presence of 
several metal lines besides iron. In particular, a 6.4 keV fluorescent 
K$\alpha$ component appears superposed to the 6.7 keV emission feature.

\subsection{The Gas Distribution}

The structure of the interstellar medium (ISM) 
in the central $10^2$ parsecs of the Galaxy has been extensively investigated
and a detailed review can be found in \markcite{gen94}Genzel, Hollenbach, \&
Townes (1994).  
Informations about the dust distribution are obtained from the re--radiation 
of UV and visible photons into the infrared continuum,
atomic and ionized components are observed directly in the 21--cm line 
while the more abundant molecular gas is sampled by the millimetre, 
submillimetre and infrared lines of trace molecules (CO, OH, HCN, ...). 
The resulting picture shows that the Galactic Center is a region 
characterized by a strong concentration of dense interstellar material, 
with an average density of gas and 
dust $10^2-10^5$ times higher than in the rest of the Galaxy. 
The central 5 parsecs contain a circum--nuclear 
disk (CND) of orbiting filaments and streamers with its inner edge at 
$\sim 1.5$ pc from the center. This disk is probably fed by the infalling 
gas from denser molecular clouds at $\rsf\gtrsim 10$ pc and drops streamers in 
the 
central region. The inner region ($\rsf\lesssim 1.5$ 
parsecs: the central cavity 
and the mini--spiral) are comparatively devoid of material and the 
average gas density is, at least, one order af magnitude lower than in the 
CND. On larger scales, surveys of 2.6 mm CO and 
far--infrared dust emission with {\it IRAS\/}
(\markcite{dam87}Dame {\it et al.\/} 1987, \markcite{db88}Deul \& Burton 1988)
show that $\sim 10^8 \, \MS$ of gas  
($\sim 10\%$ of the total Galactic ISM) are contained 
in the inner few hundred parsecs. Using a new CO--to--H$_2$ conversion factor,
\markcite{sof95a}Sofue (1995a) estimated a value for the gas mass of 
$\sim 4.6 \times 10^7 \, \MS$ within 
$\di\sim 150$ pc, where $\di$ is the distance from the center in the plane of 
the sky. The corresponding average gas density turns out to be 
$\sim 100$ cm$^{-3}$, in agreement with the value reported 
by \markcite{gen94}Genzel, Hollenbach, \& Townes (1994) for the inner 100 pc.
The total mass of gas and dust is only 1--10\% 
of the stellar mass and does not contribute significantly to the 
Galactic gravitational potential. The dust comprises only about 1\% of the 
ISM mass which is mainly in the form of molecular, atomic and ionized gas.
A fraction of the interstellar material is organized in dense molecular 
clouds and macro--structures: within $\sim
500$ pc the filling factor is $\sim 10\%$ for clouds with $n \sim 10^3$
cm$^{-3}$ and $\sim 0.05 \%$ for 
clouds with $n\sim 10^5$ cm$^{-3}$ (see, e.g., \markcite{cm93}Campana \& 
Mereghetti 1993 for a discussion). 
As recently discussed by \markcite{sof95a}Sofue (1995a), the gas distribution 
in
the central $\sim 1$ kpc region is dominated by a rotating ring with $n \approx 
10^3$ cm$^{-3}$ located at $\sim 120$ pc while the gas density 
drops by one order of magnitude outside. An expanding, roughly spheroidal, 
molecular shell is also observed at $\rsf \sim 180$ pc, $|z| \lesssim 50$ pc, 
with 
the gas mainly concentrated at intermediate latitudes (\markcite{sof95b}Sofue 
1995b).

Using {\it IRAS\/} observations of the $3^\circ \times 2^\circ$ region around 
Sgr A, \markcite{cl89}Cox \& Laureijs (1989) proposed that the dependence of 
the gas mass on the projected distance from the Galactic Center is 
\begin{equation}
m_{gas} (< \di) \sim 2 \times 10^3 \di^{1.8} \MS \, .
\end{equation}
Such distribution holds approximately up to $300$ pc from the center and 
implies, on such scales, a nearly constant surface density. As discussed
by \markcite{cl89}Cox \& Laureijs, this
result is very sensitive to the assumed dust temperature
and the value of the total gas mass is correct to about 30 \% within
300 pc (50 \% at larger distances).  
Clearly, the averaged volume density profile can be derived only by 
de--projection and any distribution matching the constraint of 
constant surface density is a priori acceptable. In particular, if the
extension of the gas along the line of sight is roughly constant when $\di$
varies, it is not unreasonable to assume an homogeneous gas distribution 
with an averaged, constant value for $n$. Within 225 pc, 
the averaged value of $n$ derived from {\it IRAS\/} data is $\sim 60$
cm$^{-3}$, assuming an elliptical gas distribution with an axial ratio 
0.7 and using a value of $3.6 \times 10^7 \MS$ for the total mass 
(\markcite{cl89}Cox \& Laureijs). However, in the inner $\sim 100$ pc this 
value is probably an underestimate (see the preceeding discussion) and 
$n = 10^2$ cm$^{-3}$ seems to be closer to 
observations. In the following sections we will focus our attention  
on the contribution of accreting ONSs to the diffuse emission from the Galactic 
Center. The accretion rate will be estimated assuming  
an homogenous distribution of the ISM with 
\begin{equation}
n =  100 \, {\rm cm^{-3}}  \qquad \di< 100\, {\rm pc}
\end{equation}

Outside this region, in order to correct the X--ray emission for 
interstellar absorption, we need an estimate of the gas density 
in the Galaxy. This is known to be of order unit and is a parameter in 
our model, adjusted in such a way to give a total column density, integrated
over $8.5$ kpc at $z=0$, always greater that $N_H = 2 \times 10^{23}$ 
cm$^{-2}$. As discussed by \markcite{pt94}Predehl \& Tr\"umper (1994), 
such a large value of $N_H$ appears compelling if the 
deficit of X--rays sources observed by {\it ROSAT\/} toward Sgr~A$^*$ has to be 
accounted for.

\subsection{The Star Distribution}

Both observations and theory (see e.g. \markcite{bai80}Bailey 1980;
\markcite{all83}Allen, Hyland, \& Hillier 1983; \markcite{san89}Sanders 1989;
\markcite{mor93}Morris 1993) provide convincing evidence that in a very 
large range of Galactocentric distances, from few arcseconds to few degrees 
from the center, the star volume density scales approximately as $\rsf^{-2}$
($\rsf$ is the galactocentric spherical radius)
in agreement with the prediction of the isothermal cluster model for the region 
outside the core radius $a$. The value of the core radius derived 
from observations ranges between $\sim 0.04$ and $\sim 0.8$ pc (see 
e.g. \markcite{mor93}Morris 1993; \markcite{gen94}Genzel, Hollenbach, \&
Townes 1994 and references therein).  
Within the core radius, data derived from the dynamics of both the
stellar and gas components agree quite well in suggesting that 
an additional dark mass of $\sim 3 \times 10^6 \, \MS$ is present in the
central region. This mass could be in the form of either
a single, massive black hole or, if the central density is not in excess of 
$10^8 \MS$ pc$^{-3}$, a cluster of many compact remnants 
(\markcite{mor93}Morris 1993; \markcite{gen94}Genzel, Hollenbach, \&
Townes  1994). 
According to this picture, the mass volume density of the stellar component
is given by 
\begin{equation}
\rho (\rsf) = \rho_c \rsf^{-1.8} \, \MS{\rm pc}^{-3}\, , 
\end{equation}
where $\rho_c$ is the central density which can be inferred from the estimated
mass inside the core radius; typical values for $\rho_c$ are a few $10^5\MS$
pc$^{-3}$ (\markcite{san89}Sanders 1989; \markcite{mor93}Morris 1993). 
In the following we assume that this profile adequately describes
the stellar distribution in the region beyond 1 pc from the Galactic Center
and we take $\rho _c = 4 \times 10^5\MS$ pc$^{-3}$. 

The fraction $f$
of the total mass comprised in neutron star is a free parameter of our model
and we assume that all neutron stars have the same mass, $M_{ns} = 1.4 \MS$. 
For a Salpeter initial mass function (IMF) 
\begin{equation}
N(M) \propto M^{-\alpha}
\end{equation}
with $\alpha = 2.35$ and assuming that all stars with initial masses 
between $\sim 10 \MS $ and $\sim 50 \MS$ have left neutron stars, 
$f$ turns out to be $\sim 1$ \%.   With a fixed $\alpha$, 
the previous formula can be assumed to describe the IMF of an 
already formed galaxy for masses in a given range, so  
possible differences due to the evolution of the bulge in the early 
Galactic history are ignored.    
However, as suggested by \markcite{mor93}Morris (1993), the initial mass 
function for 
stars formed in the Galactic Center may be quite different from that 
averaged over the disk of the Galaxy, due to the different physical conditions 
in the star--forming clouds. In particular, throughout most of the central 
region, strong tidal forces increase  
the value of the limiting density for a cloud to become 
self--gravitating and this, in turn, acts toward inhibiting 
star formation. This implies also that close to the center self--gravitating 
clouds are more and more unlikely to be 
present and that those in which the density is above the tidal limit 
are presumably supported by the strong magnetic pressure. Large internal 
velocity dispersions and higher temperatures within the 
clouds tend to increase the 
Jeans mass up to an order of magnitude with respect to the typical value in 
the rest of the Galaxy. In the inner few parsecs 
the most effective way of forming stars seems to be external 
cloud compression due 
to collisions with other clouds and/or with shock fronts produced by 
supernova explosions and 
other forms of nuclear activity. During the compression the gas 
temperature and the value of the Jeans mass are increased.  
Basing on 
these arguments \markcite{lkt82}Loose, Kr\"ugel, \& Tutukov (1982) 
argued that the mass spectrum 
in the Galactic Center is probably skewed toward higher masses, 
with a lower cutoff at 
$M_{min} \sim 1-3 \MS$ ($M_{min} = 0.08 \MS$ is often 
assumed in the disk). 
In addition, 
the value of $\alpha$ itself 
turns out to be smaller than in the disk, so 
$f=0.01$ should be regarded as a lower limit. 
Corrections due to the effects of dynamical friction over the lifetime of the 
Galaxy have been considered by \markcite{mor93}Morris (1993), but they have 
not been included 
here since they are more relevant for black holes due to their larger masses. 
Black hole remnants are expected to dominate the central cluster and their 
total mass is a large fraction of the dynamically inferred one. On the other 
hand, since mass segregation is not expected to modify 
strongly the distribution of neutron stars remnants, 
in the following we will use a number density of neutron stars given,
outside the core radius, by
\begin{equation}
n(\rsf) = f {\rho_c \over M_{ns} } \rsf^{-1.8}\, {\rm pc^{-3}}\, . 
\end{equation}

Since the total luminosity emitted in the accretion process depends on both 
the gas density and the star velocity, the velocity 
distribution $f(v)$ of neutron stars is needed in order to
estimate the cumulative emission associated with the NS population.
The velocity distribution of Galactic ONSs can be 
obtained assuming a suitable distribution of NSs at birth in phase--space 
and following the evolution of a large number of orbits in 
the Galactic gravitational potential (see e.g. \markcite{bm93}Blaes \& Madau 
1993, \markcite{zan95}Zane {\it et al.\/} 1995 and references therein). 
This kind of approach proved very useful in calculating $f(v)$ on large 
scales but, due to the particular conditions of the Galactic 
Center, results obtained from a detailed analysis are not 
necessarily more correct than a simpler estimate.
As discussed by \markcite{sel90}Sellgren {\it et al.\/} (1990, see also 
\markcite{gen94}Genzel, Hollenbach, \&
Townes 1994), the velocity dispersion $\sigma_v$ for low--mass stars in
the central region is nearly 
constant between 0.6 pc and 100 pc, with  $\sigma_v \sim 75$ km/s.  
At very small $\rsf$ observational data indicate that $\sigma_v$ increases 
up to $\sim 125$ km/s. Here we assume that NS remnants were able to reach 
energy equipartition with the field stars, establishing a Maxwellian 
distribution
\begin{equation}
f (v) = { x^2 \over v_0} \exp \left ({- 3x^2/2} \right ) \, ,
\end{equation}
where $x = v/v_0$, $v_0 = \sigma_v$ and typical values for $\sigma_v$ are in 
the range $75-125$ km/s. For a typical mass of $\sim 1 \MS$, it can be easily 
shown that the relaxation time $t_{relax}$ (see e.g. \markcite{bt87} Binney \& 
Tremaine 1987) is $\lesssim 10^{10}$ yr in the inner $\sim 10$ pc. 
Since $t_{relax}$ is a measure of the time required for deviations
from a Maxwellian distribution to be significantly decreased, in the
innermost region our hypotesis is justified. 
Energy equipartition between 
populations of different mass is reached in about the same time. In addition, 
the evaporation time turns out to be $\gtrsim 10^{10}$ yr for $r\lesssim 10$ 
pc, so the irreversible leakage of stars from the system due to stellar 
encounters is negligible over the Galaxy lifetime. A similar approach was 
used by \markcite{mor93} Morris (1993) for the velocity distribution of the 
central cluster of black hole remnants. Extending the Maxwellian
assumption up to $\sim 100$ pc is motivated on the basis of observational data.
The total mass scales approximately as $r$ here and, as
previously discussed, this is just what is expected in an isothermal
cluster model. Actually, a lowered Maxwellian distribution is
probably more reasonable, but the correction in the high energy tail 
(above $\sim v_{esc}\sim 2\sigma_v$)
is not very important as far as the fraction af accreting ONSs is
concerned, and will be not included in our calculations.

\section{Results}

In this section we estimate the contribution  
of the neutron star population to the diffuse emission near the Galactic 
Center, using the NS and gas distributions
considered in the previous sections. For a general discussion of neutron
stars accreting the ISM we refer to \markcite{tc91}Treves \& Colpi (1991)
and \markcite{bm93} Blaes \& Madau (1993).

The spectral properties of 
the radiation emitted by accreting ONSs have been recently investigated by 
\markcite{zam95}Zampieri {\it et al.} (1995). 
They have shown that, even if the atmosphere is in LTE, the emergent 
spectrum is harder than a blackbody at the neutron star effective temperature 
and the hardening ratio, typically $\sim 1.5-3$, increases with 
decreasing luminosity. This spectral hardening, produced by
the decoupling of different frequencies at different depths, is present 
independently of the assumed geometry of the accretion flow. If
ONSs retain a relic magnetic field which channels accretion onto the 
polar caps, the spectrum is harder because of the reduced emitting area
(see e.g. \markcite{tc91}Treves \& Colpi 1991) 
and the two effects add together, making
the emission of medium X--rays ($\sim 4-5$ keV) possible in the
dense region of the Galactic Center.  In the following we assume that the
emitted spectrum is that calculated by Zampieri {\it et al.\/} and that ONSs
have a relic magnetic field $B = 10^9-10^{10}$ G. Note that in this case the
luminosity released in the cyclotron line is negligible (\markcite{nel95}Nelson
{\it et al.\/} 1995).

The monochromatic flux emitted by ONSs in a region of volume $V$ centered
at $\rsf=0$ can be calculated from the expression
\begin{eqnarray}
F_\nu = \int_V n \left ( \rsf \right ) \di d \di dz d \phi \times \nonumber \\
\int_0^{\infty} f \left ( v \right ) { \tilde L_\nu \over 4 \pi R^2} dv\quad
{\rm erg \, cm^{-2} \, s^{-1}}  \, ,
\end{eqnarray}
where $\tilde L_\nu$ is the monochromatic luminosity at the source 
corrected for the interstellar absorption.
In the previous expression the volume element is expressed 
in terms of the cylindrical coordinates $(\di, z, \phi)$: $z$ is along
the line of sight and $\di$ has been defined in section 2.
$R_0=8.5$ kpc and $R = (R_0^2 + \di^2 - 2 R_0 z + z^2)^{1/2}$ are 
the distance of the Galactic Center and of the star from the Sun. The integral 
in equation (7) has been evaluated for $|z|\leq z_{max}$ and for different
values of $\di$. {\it IRAS\/} data discussed in section 2 show that the dense 
material
extends for a few hundred parsecs around the center, so we assume 
$z_{max}= 300$ pc.
The unabsorbed monochromatic flux at the source depends on the accretion rate 
which, in turn, is a function of $v$ and $n$. To avoid the direct calculation 
of the spectrum for each value of the parameters within the multiple integral, 
we have computed a set of models for different values of $\dot M$. The spectral
distribution corresponding to any given pair of values of $v$ and $n$ is then
obtained by spline interpolation.
The total emitted flux and the total flux for square degree in a fixed 
energy band are calculated integrating equation (7) over frequencies.
All multiple integrals have been evaluated numerically 
and the cross sections for the interstellar 
absorption has been taken from \markcite{mmc83}Morrison \& McCammon (1983).
 
The region within $\sim 1$ pc from the center (central cavity), where  
our assumed star and gas distributions cease to be valid,
has been excluded from the integration domain in all models. The X--ray
emission from accretion onto collapsed objects in the cavity under hypotheses
similar to our has been recently estimated by \markcite{hal96}Haller {\it et
al.\/} (1996) and turns out to be completely negligible.

Results of model calculations are compared with {\it Granat\/} ART--P
observations. 
As discussed in subsection 2.1, the diffuse emission detected by {\it Granat\/}
in the 2.5--8.5 keV band comes from a roughly elliptical region of 
$\sim 1.18$ deg$^2$ around Sgr A$^*$. The values of the observed total flux 
reported by \markcite{smp93}Sunyaev, 
Markevitch, \& Pavlinsky (1993) are $4.8 \pm 1 \times 
10^{-10}$ ergs~cm$^{-2}$~s$^{-1}$ and $4.6 \pm 1.3 \times 10^{-10}$
ergs~cm$^{-2}$~s$^{-1}$ in the two energy bands 2.5--8.5 keV and 
8.5--22 keV. For the sake of simplicity, we approximate the total ellipse 
with a circle of the same area, centered at $\rsf=0$.
We then calculate the expected 
X--ray flux in the two spectral bands, varying $B$, $v_0$ and $N_H$. 
The fraction $f$ of neutron stars is always taken equal to $0.01$;
obviously fluxes scale linearly with $f$.
Results are reported in table 1. As can be seen, the flux emitted above 8.5 keV
never exceeds a few percent of the observed one, so accreting ONSs can not
be directly responsible for the detected hard X--ray emission. On the 
contrary, the contribution of ONSs in the lower energy band can be substantial.
For the two values of $B$ we have considered, the expected emission 
ranges from a minimum of 0.07 up to $\sim 1.6$ the {\it Granat} flux in the
2.5--8.5 keV band. Clearly, this kind of calculation is necessarily influenced 
by the uncertainties in the assumed ISM and ONSs distributions. 
However, allowing for different values of the ONSs mean velocity and of the 
total column density, computed values are always comparable with observational 
data. In addition, while it could be not unreasonable to 
assume higher mean velocities for the NSs population (see e.g. 
\markcite{ll94}Lyne \& 
Lorimer 1994; \markcite{hal96}Haller {\it et al.\/} 1996), 
the range of column densities we 
have explored corresponds to the higher values for the interstellar absorption 
toward the Galactic Center reported in the literature (see e.g. 
\markcite{tho95}Thomas {\it 
et al.\/} 1996). Moreover the fraction $f$ of neutron star has been fixed to 
a rather conservative (low) value. All these considerations 
suggest that the ONSs contribution to the diffuse X--ray emission from 
the Galactic Center can be substantial in the low energy 
component.

We note that the ONSs integrated luminosity is typically $\gtrsim 10^{38}$ 
erg~s$^{-1}$ and that most of the radiation is in soft X--rays. This may prove 
of importance in the discussion of the heating and dynamics of the ISM in
the central region of the Galaxy.

The spectrum of the diffuse emission from both the total $1^\circ \times 
1.5^\circ$ ellipse and a smaller circle 30$^\prime$ in diameter has been 
obtained by \markcite{msp93}Markevitch, Sunyaev, \& Pavlinsky (1993). 
For the sake of comparison, we have calculated the monochromatic flux from
these two regions using equation (7). Results are 
presented in figures 1 and 2 for three sets of parameter values.
Crosses in figures 1 and 2 are the {\it Granat\/} data with their error bars. 
As can be seen, the observed spectral distribution is flatter than the 
predicted one, but the spectral shape is reasonably reproduced up to 6--7 keV. 
Other physical processes should be invoked to explain the harder emission.

We have compared the radial distribution of the emitted
flux with available observational data. 
From the central 30$^\prime$ 
region we expect approximately 
34\% of the total flux. This fraction should be compared with the 
$\sim 25\%$ deduced from {\it Granat\/} data. The two values are in rough 
agreement although the radial distribution of the diffuse 
emission predicted by our simplified model appears more concentrated 
toward the center than the observed one.    
Figure 3 shows the contour levels of the X--ray flux per square degree, 
normalized to the total, in the 2.5--5 keV band.

As a final point, we note that accretion onto ONSs produces an 
extremely smooth X--ray source in our model. In fact, the neutron star with 
the highest accretion rate contributes only to $3 \times 10 ^{-4}$ of the 
total X--ray luminosity. 

\section{Discussion}

In this paper we have considered the possibility that the 2.5--7 keV
component of 
the diffuse X--ray source observed in the direction of the Galactic Center 
is due to the 
unresolved emission from old neutron stars accreting the interstellar 
medium near the Galactic Center. Due to the large gas and stellar densities,
the central region of the Galaxy appears, in fact, a very promising site 
to detect the overall emission from accreting ONSs. 
We have shown that a sizable fraction, possibly most, of the 2.5--7 keV 
emission observed in the Galactic Center may be explained in terms of 
an old neutron star population accreting the dense ISM. Because of the 
intrinsic  
uncertainties in our assumptions, mainly in the value of the neutron star
relic magnetic field and in the star velocity distribution, it is not possible
to assess firmly that ONSs are responsible for the extended X--ray emission 
in this band. Moreover, accretion itself onto magnetized neutron stars could
be questionable. Although the propeller effect is not a seriuos threat 
if $B\sim 10^9$ G (see \markcite{bm93} Blaes, \& Madau 1993; \markcite{tcl93} 
Treves, Colpi, \& Lipunov 1993), preheating of the infalling gas due to the 
emitted radiation may inhibit accretion, as recently pointed out by
\markcite{bwm95} Blaes, Warren, \& Madau (1995).
Modulo these caveats, our calculations show that the observed intensity, 
spectral shape and flux radial dependence are substantially well reproduced
for different values of the free parameters of the model.  
Above $\sim 6-7$ keV the integrated ONSs spectrum gives only a marginal
contribution to the observed flux. 

The calculated continuum is rather flat and
then drops sharply above $\sim 6$ keV, so we do not expect that ONSs
emission could provide an important photoionization source to produce 
ultimately iron lines. However, our synthetic spectra were computed 
considering only pure hydrogen atmospheres around accreting neutron stars.
A definite assessment about the presence of K$\alpha$ lines in the emerging
spectrum would require the extension of our model to include line processes
and a more realistic chemical composition for the accreting gas.
If accretion onto neutron stars with a magnetic field as high as 
$\sim 10^{12}$ G is possible (which is dubious, see, e.g., \markcite{tcl93} 
Treves, Colpi, \& Lipunov 1993), about 10 \% of the flux is emitted in the
cyclotron line at $E_B = 11.6 B_{12}$ keV (\markcite{nel95}Nelson 
{\it et 
al.\/} 1995) and the excitation of the Fe lines could be substantial. 

\markcite{smp93}Sunyaev, Markevitch, \& Pavlinsky (1993) 
noticed some correlation 
between the distribution of giant molecular clouds and the X--ray brightness 
and interpreted it as a signature of the reprocessing of medium
energy photons by the clouds themselves. Such a correlation is naturally 
explained in our scenario, since GMCs provide regions of very high density in
which the accretion rate can be substantially higher than our present estimate,
based on $n=100$ cm$^{-3}$. This implies also harder spectra, so that the 
emission of hard photons, $\sim 10$ keV, is possible for ONSs accreting
in clouds with $n\sim 10^4$ cm$^{-3}$. On this regard, see the discussion of 
\markcite{cm93}Campana \& Mereghetti (1993) on the source 1740.7-2942.

The correlation between X--ray emission and the ISM distribution
may become the basic probe to test the real relevance 
of ONSs accretion in explaining the diffuse X--ray source in the Galactic
Center. AXAF and JET--X  will have the required sensitivity and space 
resolution in medium energy X--rays to shed light on this important issue.

\clearpage
%

\begin{deluxetable}{ccccc}
\tablecolumns{5}
\tablenum{1}
\tablecaption{Computed fluxes for different parameter values\label{tab1}}
\tablehead{
\colhead{$v_0$} & 
\colhead{$N_H $}& 
\colhead{$B$} &
\colhead{$F_{[2.5-8.5]}$} &
\colhead{$F_{[8.5-22]}$} \\
\colhead{km s$^{-1}$} & 
\colhead{$10^{23}$ cm$^{-2}$} & 
\colhead{G} &
\colhead{$ 10^{-10}$ erg cm$^{-2}$ s$^{-1}$} & 
\colhead{$ 10^{-10}$ erg cm$^{-2}$ s$^{-1}$}}
\startdata
   75 &  2    &$10^{9}$ & 7.11 &  0.071 \nl
   75 &  3    &$10^{9}$ & 3.90 &  0.064 \nl
   75 &  4    &$10^{9}$ & 2.44 &  0.057 \nl
  100 &  2    &$10^{9}$ & 3.18 &  0.031 \nl
  100 &  2.5  &$10^{9}$ & 2.30 &  0.030 \nl
  100 &  3    &$10^{9}$ & 1.74 &  0.028 \nl
  100 &  3.5  &$10^{9}$ & 1.36 &  0.027 \nl
  100 &  4    &$10^{9}$ & 1.09 &  0.025 \nl
  125 &  2    &$10^{9}$ & 1.68 &  0.017 \nl
  125 &  3    &$10^{9}$ & 0.92 &  0.015 \nl
  125 &  4    &$10^{9}$ & 0.57 &  0.013 \nl
  150 &  2    &$10^{9}$ & 0.99 &  0.010 \nl
  100 &  2    &$10^{10}$& 7.80 &  0.151 \nl
  100 &  3    &$10^{10}$& 4.61 &  0.135 \nl
  100 &  4    &$10^{10}$& 3.03 &  0.121 \nl
  200 &  2    &$10^{10}$& 1.05 &  0.020 \nl
  300 &  2    &$10^{10}$& 0.32 &  0.006 \nl
\tablecomments{
$ f = 10^{-2}$, $ a = 1$ pc, 
$ z_{max} = 300 $ pc}
\enddata
\end{deluxetable}

\clearpage

\figcaption[fig1.eps]{{\it Granat\/} ART--P data (Markevitch, Sunyaev, \& 
Pavlinski 1995, crosses) and 
calculated X--ray spectra from a region of 1.18 deg$^2$ for:
a) $f=0.01$, $v_0=75$ km/s and $N_H=3\times 10^{23}$ cm$^{-2}$ (solid line);
b) $f=0.02$, $v_0=100$ km/s and $N_H=3\times 10^{23}$ cm$^{-2}$ (dashed 
line); c) $f=0.015$, $v_0=100$ km/s and $N_H=2.5\times 10^{23}$ cm$^{-2}$ 
(dash--dotted line). For all models $B=10^9$ G.\label{fig1}}

\figcaption[fig2.eps]{Same as in figure 1 for the central region $30\arcmin$ 
in diameter.\label{fig2}}

\figcaption[fig3.eps]{Contour levels of the X--ray flux/deg$^2$ in the 
2.5--5 keV band. Parameters values are $B = 10^9$ G, 
$f=0.01$, $v_0=75$ km/s and $N_H=3\times 10^{23}$ cm$^{-2}$.\label{fig3}}


\clearpage

\begin{references}

\reference{all83} Allen, D.A., Hyland, A.R., \& Hillier, D.J. 1983, \mnras, 
204, 1145
\reference{bai80} Bailey, M.E. 1980, \mnras, 190, 217
\reference{bzc96} Belloni, T., Zampieri, L., \& Campana, S. 1996, \aap, 
in press
\reference{bt87} Binney, J., \& Tremaine, S. 1987, Galactic Dynamics (Princeton:
Princeton University Press)
\reference{bm93} Blaes, O., \& Madau, P. 1993, \apj, 403, 690
\reference{bwm95} Blaes, O., Warren, O., \& Madau, P. 1995, \apj, 454, 370 
\reference{cm93} Campana, S., \& Mereghetti, S. 1993, \apjl, 413, L89
\reference{cct93} Colpi, M., Campana, S., \& Treves, A. 1993, \aap, 278, 161
\reference{cl89} Cox, P., \& Laureijs, R. 1989, in IAU Symp. 136, The Center 
of the Galaxy, ed. M. Morris (Dordrecht: Kluwer), 121
\reference{dkh94} Danner, R., Kulkarni, S.R., \& Hasinger, G. 1994, poster 
paper presented at the 17th Texas Symposium on Relativistic Astrophysics and 
Cosmology, M\"unchen, Germany
\reference{dam87} Dame T.M., et al. 1987, \apj, 322, 706
\reference{db88} Deul, E.R., \& Burton, W.B. 1988, Mapping the Sky: Past 
Heritage and Future Directions, ed. S. Debarat et al. (Dordrecht: Kluwer)
\reference{gen94} Genzel, R., Hollenbach, D., \& Townes, C.H. 1994, Rep. Prog. 
Phys., 57, 417
\reference{hal96} Haller, J.W., et al. 1996, \apj, 456, 194
\reference{kaw88} Kawai, N., et al. 1988, \apj, 330, 130
\reference{kel71} Kellogg, E., et al. 1971, \apjl, 169, L99
\reference{koy89} Koyama, K., et al. 1989, \nat, 339, 603
\reference{koy96} Koyama, K., et al. 1996, in Proceedings of the conference
R\"ontgenstrahlung from the Universe, eds. H.U. Zimmermann, J.E. Tr\"umper
\& H. Yorke, MPE Report 263, 315
\reference{lkt82} Loose, H.H., Kr\"ugel, E., \& Tutukov, A. 1982, \aap, 105, 342
\reference{ll94} Lyne, A.G., \& Lorimer, D.R. 1994, \nat, 369, 127
\reference{mb94} Madau, P., \& Blaes, O. 1994, \apj, 423, 748
\reference{mg95} Maoz, E., \& Grindlay, J.E. 1995, \apj, 444, 183
\reference{mtt73} Maraschi L., Treves A., \& Tarenghi M. 1973, \aap, 25, 
153 
\reference{msp93} Markevitch, M., Sunyaev, R.A., \& Pavlinsky, M. 1993, \nat, 
364, 40
\reference{mor93} Morris, M. 1993, \apj, 408, 496
\reference{mmc93} Morrison, R., \& McCammon, D. 1983, \apj, 270, 119
\reference{nel95} Nelson, R.W., Wang, J.C.L., Salpeter, E.E., \& Wasserman, I. 
1995, \apjl, 438, L99
\reference{not93} Nottingham, M.R., et al. 1993, \aap \ Suppl., 97, 165 
\reference{ors70} Ostriker, J.P., Rees, M.J., \& Silk, J. 1970, \aplett, 6, 179
\reference{pt94} Predehl, P., \& Tr\"umper, J. 1994, \aap, 290, L29
\reference{san89} Sanders, R.H. 1989, in IAU Symp. 136, The Center of the 
Galaxy, ed. M. Morris (Dordrecht: Kluwer), 77
\reference{sel90} Sellgren, K., et al. 1990, \apj, 359, 112
\reference{shv71} Shvartsman, V.F. 1971, Sov. Astron.--AJ, 14, 662   
\reference{ski87} Skinner, G.K., et al. 1987, \nat, 330, 544
\reference{ski89} Skinner, G.K. 1989, in IAU Symp. 136, The Center of the 
Galaxy, ed. M. Morris (Dordrecht: Kluwer), 567
\reference{sof95a} Sofue, Y. 1995a, \pasj, 47, 527
\reference{sof95b} Sofue, Y. 1995b, \pasj, 47, 551
\reference{sto95} Stocke, J.T., et al. 1995, \aj, 109, 1199
\reference{smp93} Sunyaev, R.A., Markevitch, M., \& Pavlinsky, M. 1993, \apj, 
407, 606
\reference{tho95} Thomas, B., et al. 1996, in Proceedings of the conference
R\"ontgenstrahlung from the Universe, eds. H.U. Zimmermann, J.E. Tr\"umper
\& H. Yorke, MPE Report 263
\reference{tc91} Treves, A., \& Colpi, M. 1991, \aap, 241, 107
\reference{tcl93} Treves, A., Colpi, M., \& Lipunov, V.M. 1993, \aap, 269, 319
\reference{wal96} Walter, F.M., Wolk, S.J., \& Neuh\"auser, R. 1996, \nat, 
379, 233
\reference{wat81} Watson, M.G., et al. 1981, \apj, 250, 142
\reference{yam90} Yamauchi, S., et al. 1990, \apj, 365, 532
\reference{zam95} Zampieri, L., Turolla R., Zane S., \& Treves, A. 1995, \apj, 
439, 849
\reference{zan95} Zane, S., et al. 1995, \apj, 451, 739
\reference{zan96} Zane, S., Zampieri, L., Turolla, R., \& Treves, A. 1996, 
\aap, in press
\end{references}
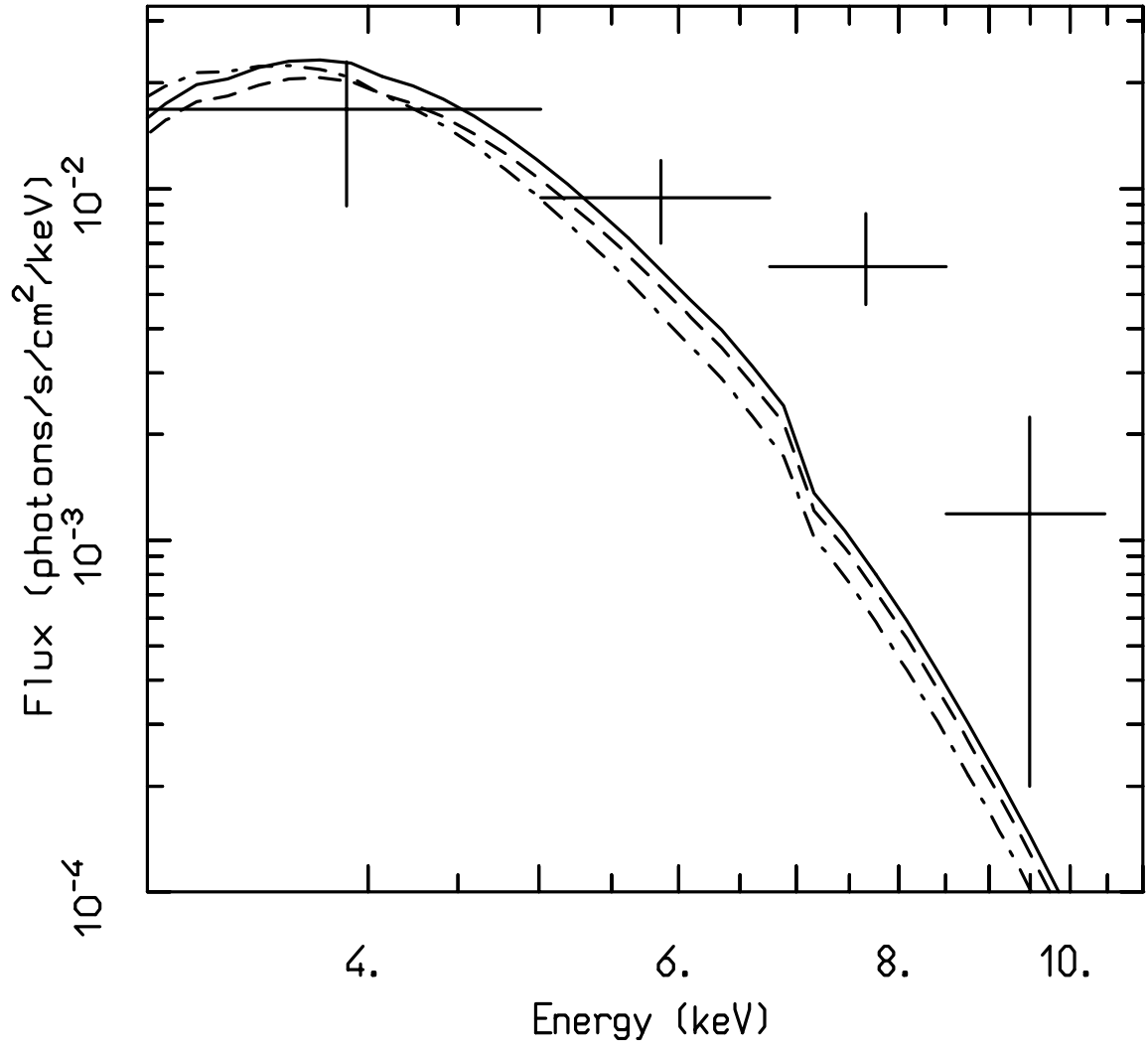
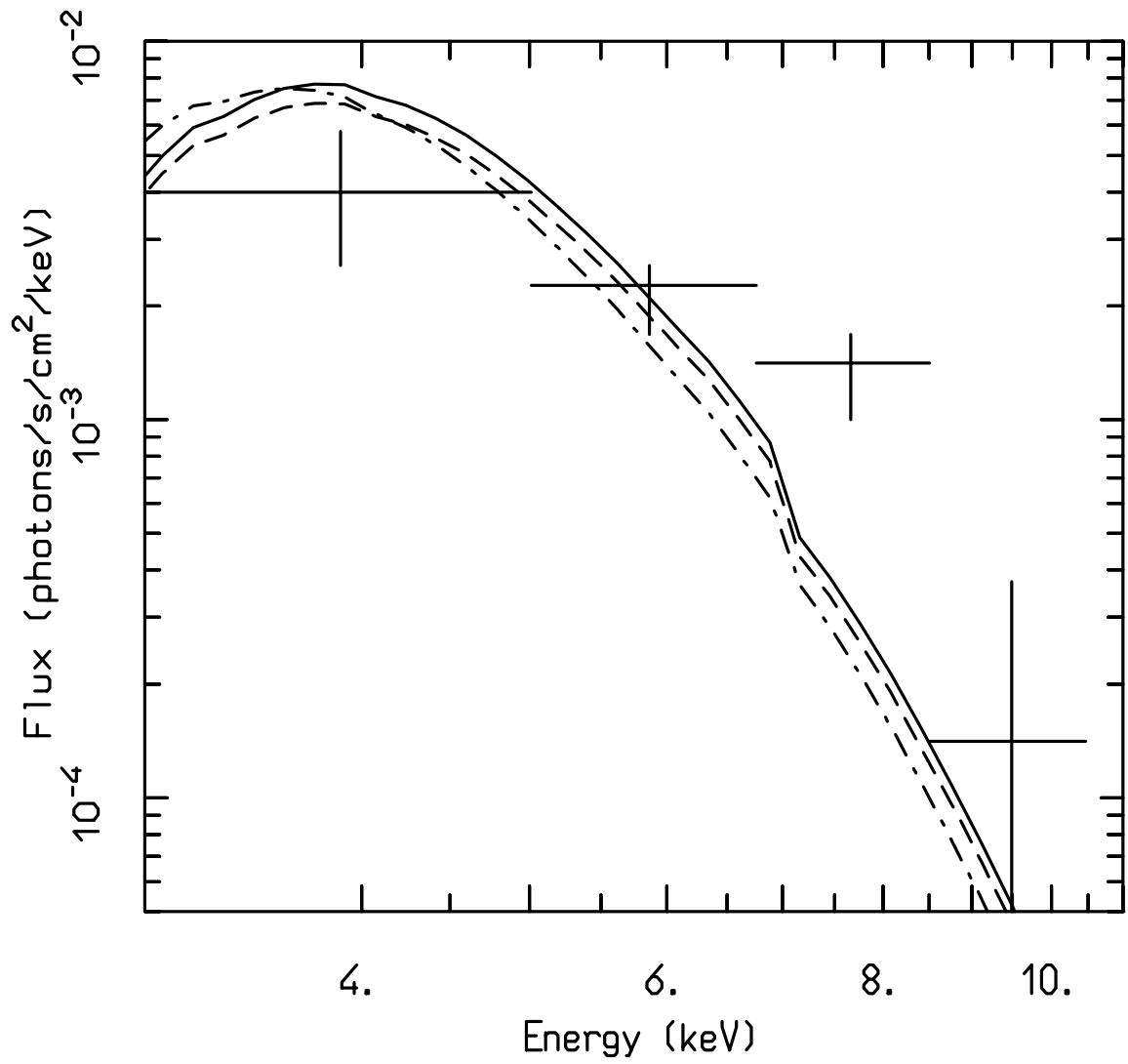
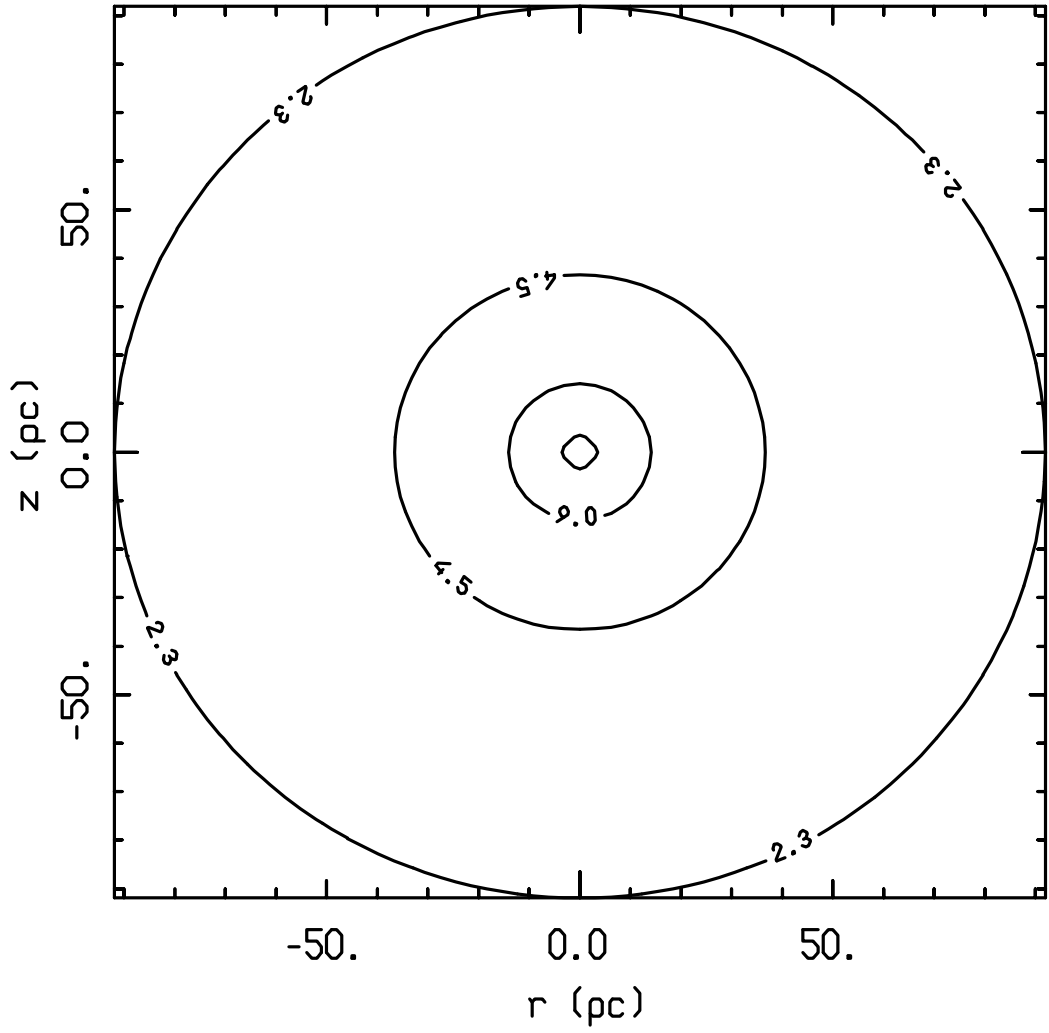
\end{document}